# Prediction and analysis of Coronavirus Disease 2019

Lin Jia[1]   Kewen Li[1, 2]   Yu Jiang[1]   Xin Guo[1]   Ting zhao[1]

[1] China university of Geosciences (Beijing), 29 Xueyuan Road, 100083 Beijing, China

[2] Stanford University, Stanford, CA 94305, USA

**Abstract**

In December 2019, a novel coronavirus was found in a seafood wholesale market in Wuhan, China. WHO officially named this coronavirus as COVID-19. Since the first patient was hospitalized on December 12, 2019, China has reported a total of 78,824 confirmed CONID-19 cases and 2,788 deaths as of February 28, 2020. Wuhan's cumulative confirmed cases and deaths accounted for 61.1% and 76.5% of the whole China mainland , making it the priority center for epidemic prevention and control. Meanwhile, 51 countries and regions outside China have reported 4,879 confirmed cases and 79 deaths as of February 28, 2020. COVID-19 epidemic does great harm to people's daily life and country's economic development. This paper adopts three kinds of mathematical models, i.e., Logistic model, Bertalanffy model and Gompertz model. The epidemic trends of SARS were first fitted and analyzed in order to prove the validity of the existing mathematical models. The results were then used to fit and analyze the situation of COVID-19. The prediction results of three different mathematical models are different for different parameters and in different regions. In general, the fitting effect of Logistic model may be the best among the three models studied in this paper, while the fitting effect of Gompertz model may be better than Bertalanffy model. According to the current trend, based on the three models, the total number of people expected to be infected is 49852-57447 in Wuhan,12972-13405 in non-Hubei areas and 80261-85140 in China respectively. The total death toll is 2502-5108 in Wuhan, 107-125 in Non-Hubei areas and 3150-6286 in China respetively. COVID-19 will be over p robably in late-April, 2020 in Wuhan and before late-March, 2020 in other areas respectively.

**Keywords**：  Mathematic model   COVID-19   epimedic prediction

## 1 Introduction

A number of unexplained pneumonia cases have successively been discovered in China since December 2019, which have been confirmed to be acute respiratory infectious diseases caused by a novel coronavirus. The outbreak of COVID-19 has experienced three stages since mid-December 2019: local outbreak, community transmission and large-scale transmission. ①Local outbreak stage: This stage mainly forms a local outbreak among the people exposed to the seafood market before the end of December 2019. Most of the cases at this stage were related to the exposure of seafood market. ② Community transmission stage: due to the spread of the epidemic, the virus spread to communities through the early-infected people, forming community transmission. Interpersonal and cluster transmission occurred in multiple communities and families in Wuhan. ③The stage of large-scale transmission of the spread of the epidemic : The epidemic rapidly expanded and spread from Hubei Province to other parts of China due to the great mobility of personnel during the Chinese Lunar New Year, while the number of COVID-19 cases in other countries gradually increased.

As of 24:00 on February 29, 2020, China has reported a total of 79,824 confirmed cases of COVID-19 and 2,870 deaths[1]. The cumulative number of confirmed cases and deaths in Wuhan accounted for 61.5% and 76.5% of the country respectively, which is the priority area for epidemic prevention and control. At the same time, countries and regions outside China reported 7,661 confirmed cases and a total of 121 deaths. Infectious diseases cause disastrous harm to human society and are one of the important factors that seriously threaten human life and health, restrict social and economic development and endanger national security and stability. The effects of economic globalization, internationalization of production, more convenient transportation, and faster human and cargo flows have created favorable conditions for the widespread spread of infectious diseases, making the spread of infectious diseases faster and wider[2]. Some infectious diseases that have occurred in recent years, such as COVID-19, SARS in 2003, influenza HIN1, H5N1, etc., have greatly affected human health and social life. How to contain the outbreak of infectious diseases and ease the spread of infectious diseases is an urgent issue facing the society at present[3].

Theoretical analysis, quantitative analysis and simulation are needed for the prediction of various infectious diseases. The above analysis cannot be carried out without models established for various infectious diseases.

Infectious disease transmission is a complicated diffusion process occurring in the crowd. Models can be established for this process to analyze and study the transmission process of infectious diseases theoretically[4], so that we can accurately predict the future development trend of infectious diseases[5]. Therefore, in order to control or reduce the harm of infectious diseases, the research and analysis of infectious disease prediction models have become a hot research topic[6].

**1.1 Traditional infectious disease prediction model**

Traditional infectious disease prediction models mainly include differential equation prediction models and time series prediction models based on statistics and random processes.

The differential equation prediction models are to establish a differential equation that can reflect the dynamic characteristics of infectious diseases according to the characteristics of population growth, the occurrence of diseases and the laws of transmission within the population. Through qualitative and quantitative analysis and numerical simulation of the model dynamics, the occurrence process of diseases is displayed, the transmission laws are revealed, the change and development trends are predicted, the causes and key factors of disease transmission are analyzed, the optimal strategies for prevention and control are sought, and the theoretical basis and quantitative basis are provided for people to make prevention and control decisions. Common models for predicting infectious disease dynamics differential equations have ordinary differential systems, which directly reflect the relationship between the instantaneous rate of change of individuals in each compartment and the corresponding time of all compartments. Partial differential system is a common model system when considering age structure. Delay differential system is a kind of differential system that appears when the structure of the stage is considered (e.g. the infected person has a definite infectious period, the latent person has a definite incubation period, the immunized person has a definite immune period, etc. The currently widely studied and applied models include SI model, SIS model, SIR model and SEIR model, etc[7]. System individuals are divided into different categories, and each category is in a state, respectively: S (Susceptible), E(Exposed), I (Infected) and R (Remove).

The classical differential equation prediction model assumes that the total number of people in a certain area is a constant, which can prompt the natural transmission process of infectious diseases, describe the evolution relationship of different types of nodes with time, and reveal the overall information transmission law. However, in practice, the population is changing over time. There will always be some form of interaction with other populations in terms of food, resources and living space. The connection between individuals is random, and the difference between spreading individuals is ignored, thus limiting the application scope of the model.

Time series prediction models, based on statistics and random processes, predict infectious diseases by analyzing one-dimensional time series of infectious disease incidence, mainly including Autoregressive Integrated Moving Average model (ARIMA), Exponential Smoothing method (ES), Grey Model (GM), Markov chain method (MC), etc. The widely used time series prediction model is ARIMA prediction model, which uses several differences to make it a stationary series, and then represent this sequence as a combination autoregression about the sequence up to a certain point in the past[8].

The infectious disease prediction model established by this method relies on curve fitting and parameter estimation of available time series data, so it is difficult to apply it to a large number of irregular data.

**1.2 Internet-based infectious disease prediction model**

Infectious disease surveillance research based on the Internet has begun to rise since the mid-1990s[9]. It can provide information services for public health management institutions, medical workers and the public. After analyzing and processing, it can provide users with early warning and situational awareness information of infectious diseases[10].

In the early research, traditional Web page web information (for example, related news topics, authoritative organizations, etc.) was the main data source. However, with the development of the Internet, research has begun to expand data sources to social media (such as Twitter, Facebook, microblog, etc.) and multimedia information in recent years. Due to the global spread of the Internet, people use Internet search engines, social networks and online map tools to track the frequency and location information of query keywords, strengthen the integration of information on social, public focus and hot issues, realize disease monitoring based on search engines and social media, and predict the incidence of infectious diseases, which can provide important reference for the decision and management of infectious disease prevention and control[11].

In theory, Internet search tracking is efficient, and can reflect the real-time status of infectious diseases. Therefore, the infectious disease prediction models based on Internet and search engine are good supplement to the traditional infectious disease prediction models[12]. U.S. scientists compared the flu estimates in different countries and regions from 2004 to 2009 with the official flu surveillance data, and found that the estimates from Google search engine were close to historical flu epidemic[13]. Jiwei et al. filtered the Twitter data stream, retained flu-related information, and tagged the information with geographic location to show where the flu-related Twitter information came from and how the information changed over a certain period of time. They counted 3.6 million flu-related Twitter messages published by about 1 million users from June 2008 to June 2010, showing that there is a highly positive correlation between Twitter's influenza information and influenza outbreak data provided by the U.S. Centers for Disease Control and Prevention[14]. In 2011, Google launched Google Dengue Trends (GDT) and in 2016, Google Flu Trends (GFT) and other tools to quantitatively track the spread trend of infectious diseases such as dengue fever and influenza in multiple regions of the world according to Google's search patterns[15].

Compared with the traditional prediction models, the Internet-based infectious disease prediction models have the advantages of real-time and fast, which can predict the incidence trend of infectious diseases as early as possible, and are suitable for data analysis of a large number of people. However, the sensitivity, spatial resolution and accuracy of its prediction need to be further improved. So Internet-based infectious disease prediction models cannot replace the traditional prediction models, and they can just be used as an extension of the traditional infectious disease prediction model[16].

This paper will use 2003 SARS data to verify three mathematical models (Logistic model, Bertalanffy model and Gompertz model) to predict the development trend of the virus, and then use these three models to fit and analyze the epidemic trend of COVID-19 in Wuhan, mainland China and non-Hubei areas, including the total number of confirmed cases, the number of deaths and the end time of the epidemic.

**1.3 Early Prediction Model of Infectious Diseases Based on Machine Learning**

In short, machine learning is to learn more useful information from a large amount of data using its own algorithm model for specific problems. Machine learning spans a variety of fields, such as medicine, computer science, statistics, engineering technology, psychology, etc[17]. For example, neural network, a relatively mature machine learning algorithm, can simulate any high-dimensional non-linear optimal mapping between input and output by imitating the processing function of the biological brain's nervous system. When faced with complex data relations, the traditional statistical method is not such effective, which may not receive accurate results as the neural network[18].

Since most new infectious diseases occurring in human beings are of animal origin (animal infectious diseases), it is an effective prerequisite to predict diseases by determining the common intrinsic characteristics of species and environmental conditions that lead to the overflow of new infections. By analyzing the intrinsic characteristics of wild species through machine learning, new reservoirs (mammals) and carriers (insects) of zoonotic diseases can be accurately predicted[19]. The overall goal of machine learning-based approach is to extend causal inference theory and machine learning to identify and quantify the most important factors that cause zoonotic disease outbreaks, and to generate visual tools to illustrate the complex causal relationships of animal infectious diseases and their correlation with zoonotic diseases[20]. However, the highly nonlinear and complex problems to be analyzed in the early prediction model of infectious diseases based on machine learning usually lead to local minima and global minima, leading to some limitations of the machine learning model.

**2 Mathematical model**

Infectious disease prediction models mainly include differential equation prediction models based on dynamics and time series prediction models based on statistics and random processes, Internet-based infectious disease prediction model and machine learning methods. Some models are too complicated and too many factors are considered, which often leads to over-fitting. In this paper, Logistic model, Bertalanffy model and Gompertz model, which are relatively simple but accord with the statistical law of epidemiology, are selected to predict the epidemic situation of COVID-19. After the model is selected, the least square method is used for curve fitting. Least square method is a mathematical optimization technique. It finds the best function match of data by minimizing the sum of squareed errors. Using the least square method, unknown data can be easily obtained, and the sum of squares of errors between these obtained data and actual data is minimized.

**2.1 Model Selection**

(1) Logistic model

Logistic model is mainly used in epidemiology. It is commonly to explore the risk factors of a certain disease, and predict the probability of occurrence of a certain disease according to the risk factors. We can roughly predict the development and transmission law of epidemiology through logistic regression analysis,.

$$Q_t = \frac{a}{1+e^{b-c(t-t_0)}} \tag{1}$$

$Q_t$ is the cumulative confirmed cases (deaths); $a$ is the predicted maximum of confirmed cases (deaths). $b$ and $c$ are fitting coefficients. $t$ is the number of days since the first case. $t_0$ is the time when the first case occurred.

(2) Bertalanffy model

Bertalanffy model is often used as a growth model. It is mainly used to study the factors that control and affect the growth. It is used to describe the growth characteristics of fish. Other species can also be used to describe the growth of animals, such as pigs, horses, cattle, sheep, etc. and other infectious diseases. The development of infectious diseases is similar to the growth of individuals and populations. In this paper, Bertalanffy model is selected to describe the spread law of infectious diseases and to study the factors that control and affect the spread of COVID-19.

$$Q_t = a(1 - e^{-b(t-t_0)})^c \tag{2}$$

$Q_t$ is the cumulative confirmed cases (deaths); $a$ is the predicted maximum of confirmed cases (deaths). $b$ and $c$ are fitting coefficients. $t$ is the number of days since the first case. $t_0$ is the time when the first case occurred.

(3) Gompertz model

The model was originally proposed by Gomperts (Gompertz,1825) as an animal population growth model to describe the extinction law of the population. The development of infectious diseases is similar to the growth of individuals and populations. In this paper, Gompertz model is selected to describe the spread law of infectious diseases and to study the factors that control and affect the spread of COVID-19.

$$Q_t = ae^{-be^{-c(t-t_0)}} \qquad (3)$$

$Q_t$ is the cumulative confirmed cases (deaths); $a$ is the predicted maximum of confirmed cases (deaths). $b$ and $c$ are fitting coefficients. $t$ is the number of days since the first case. $t_0$ is the time when the first case occurred.

**2.2 Model Evaluation**

The regression coefficient ($R^2$) is used to evaluate the fitting ability of various methods and can be obtained by the following equation.

$$R^2 = 1 - \frac{\sum(y_i - \hat{y}_i)^2}{\sum(y_i - \bar{y})^2} \qquad (4)$$

$y_i$ is the actual cumulative confirmed COVID-19 cases; $\hat{y}_i$ is the predicted cumulative confirmed COVID-19 cases; $\bar{y}$ is the average of the actual cumulative confirmed COVID-19 cases. The closer the fitting coefficient is to 1, the more accurate the prediction.

# 3 Fitting and analysis of SARS epidemic

As COVID-19 and SARS virus are both coronaviruses, the infection pattern may be similar. Firstly, we used SARS data to verify the rationality of our model.

**3.1 Number of Confirmed Cases**

The cumulative confirmed SARS data after April 21, 2003 were selected to be fitted by Gompertz model, Logistic model and Bertalanffy model. The results are shown in Figure 1. From the overall view, these three models can accurately predict the cumulative number of confirmed cases, in which Logistic model and Gompertz model are better than Bertalanffy model.

The number of confirmed SARS cases no longer increased after June 11, 2003. On June 24, 2003, WHO announced the end of the SARS epidemic, that is, the time when the cumulative number of confirmed SARS cases reached the peak value was basically the time when the epidemic ended. We used this rule to predict the end of COVID-19 epidemic[21].

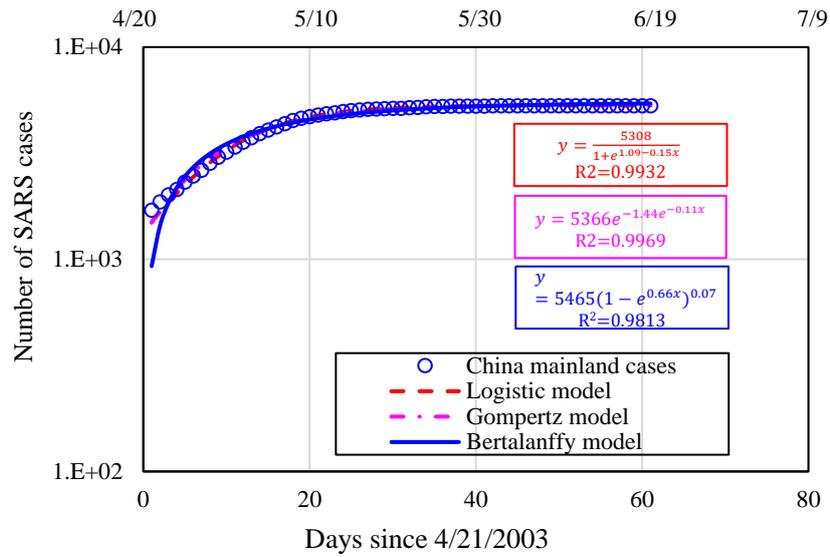

(a) Three models for predicting SARS cases since April 21, 2003

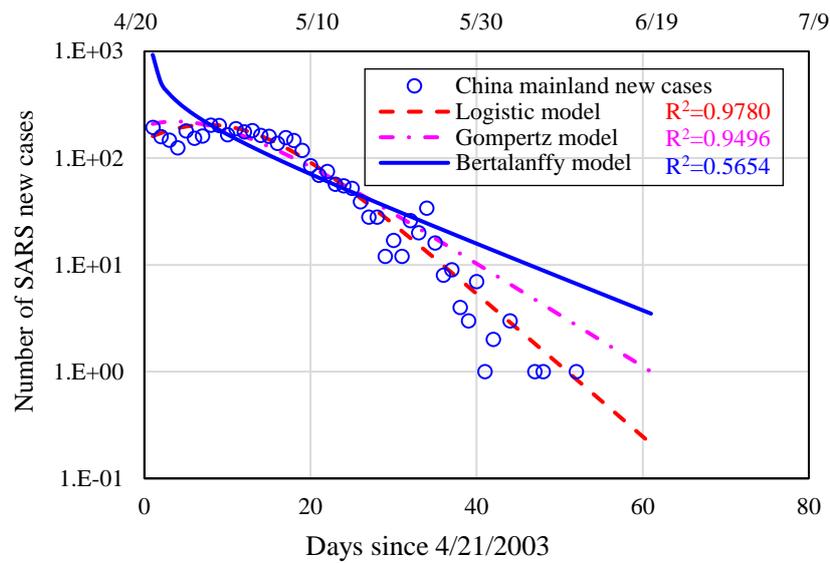

(b) Three models for predicting new confirmed SARS cases since April 21, 2003

**Figure 1 The prediction of cumulative number of confirmed SARS cases fitted by Gompertz, Logistic and Bertalanffy models**

### 3.2 Death Toll

The fitting of the death toll of SARS in 2003 is shown in Figure 2, which shows that Gompertz model and Logistic model can predict the death toll of SARS well. The good fitting of the death toll of SARS in 2003 shows that Gompertz model and Logistic model can also be used to predict that of the COVID-19 epidemic.

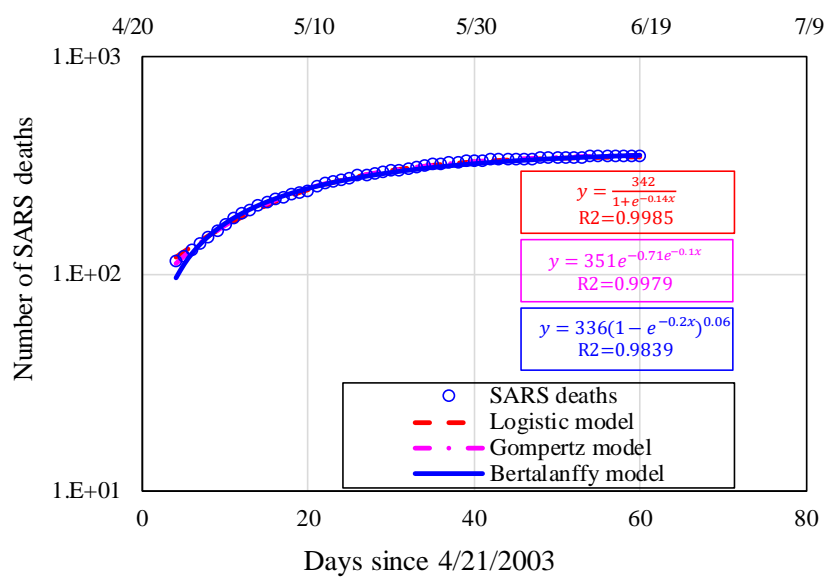

**Figure 2** The prediction of SARS death toll in mainland China fitted by Gompertz, Logistic and Bertalanffy models

## 4 Fitting and analysis of COVID-19 epidemic

### 4.1 Number of Confirmed Cases

The cumulative number of confirmed cases of the novel coronavirus (hereinafter referred to as COVID-19 ) is shown in Figure 3. The number of confirmed cases has dramatically increased in China since the first case was confirmed. The epidemic spreads to other parts of Hubei and the whole country with Wuhan as the center. Since the main confirmed cases were in Wuhan, the development trend of new confirmed cases in the whole country are basically the same with Wuhan. Judging from the prediction results, the three models can predict the epidemic situation of COVID-19 well in the later stage of the epidemic. Among them, Logistic model is better than the other two models in fitting all the data in Wuhan, while Gompertz model is better in fitting the data outside Wuhan.

Due to various reasons, It is worth noting that the number of confirmed cases suddenly increased by 13,332 on February 12, 2020.Obviously, the mutation of this data does not originate from the mechanism of the virus, so our treatment method is to remove the impact of this part of data mutation (13,332 people).Then, the impact of the sudden increase confirmed cases will be considered in the later fitting analysis.

According to the daily real-time updated data of COVID-19, we used the above three mathematical models (Logistic model, Bertalanffy model and Gompertz model) to carry out fitting analysis on the epidemic of COVID-19. The prediction results are shown in Table 1 in which a is the prediction of cumulative confirmed number (the final predicted cumulative confirmed number = a+13332); b and c are fitting coefficients; t is the number of days since the first case. $R^2(C)$ means the fitting goodness of cumulative confirmed cases, $R^2(N)$ means the fitting goodness of new confirmed cases

According to the calculation results of the three models, it is estimated that the final cumulative number of confirmed cases of COVID-19 in Wuhan is 49852-57447. Non-Hubei areas: 12972-13405; China mainland: 80261-85140, respectively.

**Table 1 The prediction epidemic results of COVID-19 in Logistic Model, Bertalanffy Model and Gompertz Model**

| Model | Parameter | Wuhan | China mianland | Non-Hubei areas |
|---|---|---|---|---|
| Logistic model | a | 36520 | 66929 | 12972 |
| | b | 5.51 | 4.98 | 4.87 |
| | c | 0.21 | 0.22 | 0.26 |
| | Cumulative number of cases | 49852 | 80261 | 12972 |
| | $R^2(C)$ | 0.9991 | 0.9993 | 0.9993 |
| | $R^2(N)$ | 0.8124 | 0.9183 | 0.9648 |
| Gompertz model | a | 42926 | 70324 | 1332 |
| | b | 17.33 | 10.98 | 15.01 |
| | c | 0.12 | 0.11 | 0.17 |
| | Cumulative number of cases | 56258 | 83656 | 13328 |
| | $R^2(C)$ | 0.999 | 0.9934 | 0.9998 |
| | $R^2(N)$ | 0.813 | 0.3372 | 0.9804 |
| Bertalanffy model | a | 44115 | 71808 | 13405 |
| | b | 14.93 | 11.5 | 13.05 |
| | c | 0.12 | 0.12 | 0.16 |
| | Cumulative number of cases | 57447 | 85140 | 13405 |
| | $R^2(C)$ | 0.9989 | 0.9993 | 0.9998 |
| | $R^2(N)$ | 0.8105 | 0.895 | 0.978 |

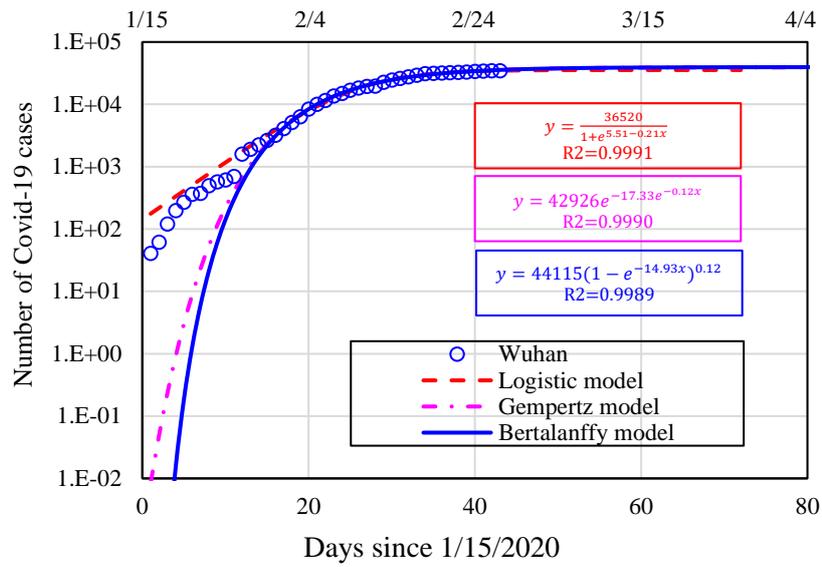

(a) Three models for predicting COVID-19 cases in Wuhan since January 15, 2020

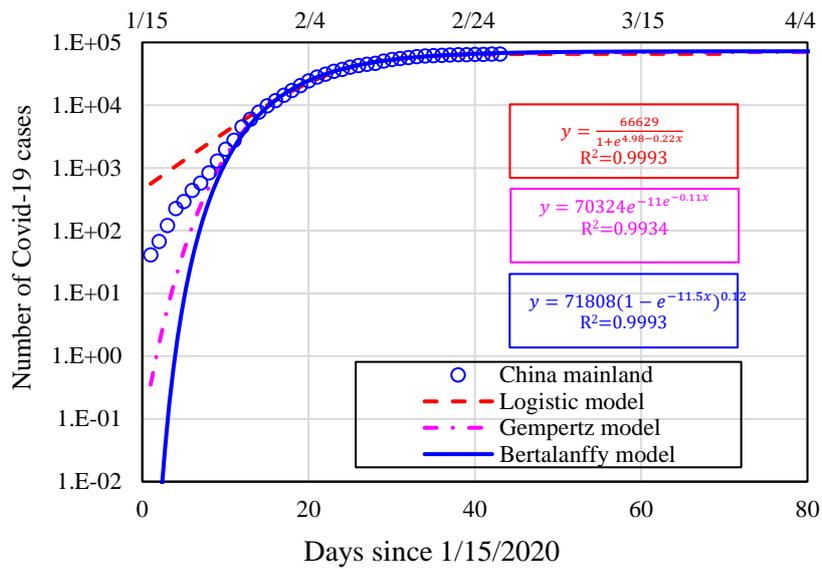

(b) Three models for predicting COVID-19 cases in China mainland since January 15, 2020.

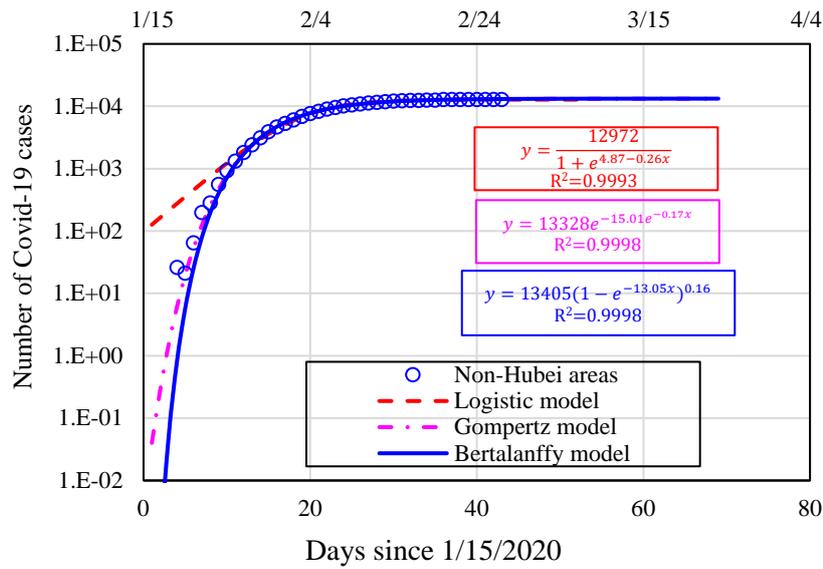

(c) Three models for predicting COVID-19 cases in non-Hubei areas since January 15,2020

**Figure 3 The prediction of cumulative number of COVID-19 cases fitted by Gompertz, Logistic and Bertalanffy models**

In order to predict the turning point, we use the above three models to compare the new confirmed cases in Wuhan, China mainland and non-Hubei areas. As can be seen in Figure 4, the turning points in W Wuhan, China mainland and non-Hubei areas are February 9, February 6 and February 2, 2020 respectively. From the results, for the prediction of newly confirmed cases, the three models can predict the COVID-19 epidemic well in the early and late stages of the epidemic. Among them, the Logistic model is better than the other two models in fitting all the data.

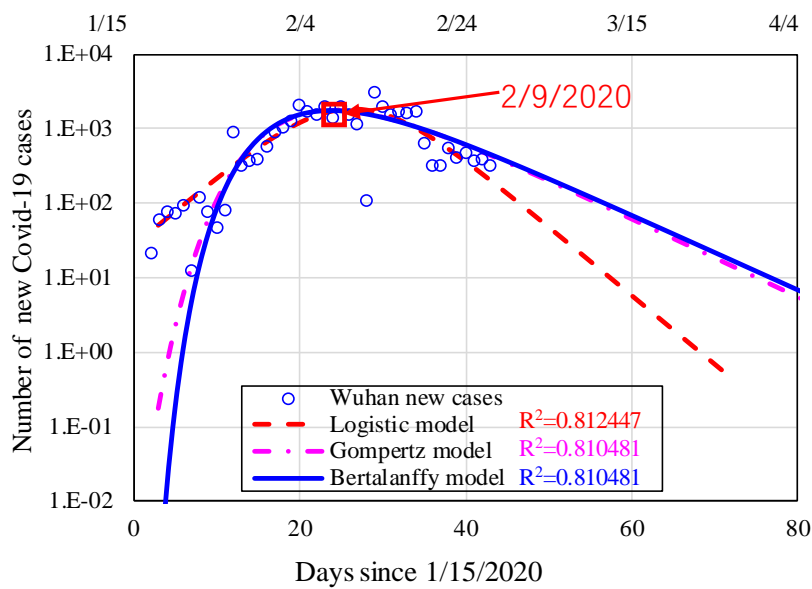

(a) Three Models for Predicting new COVID-19 Cases in Wuhan since January 15,2020

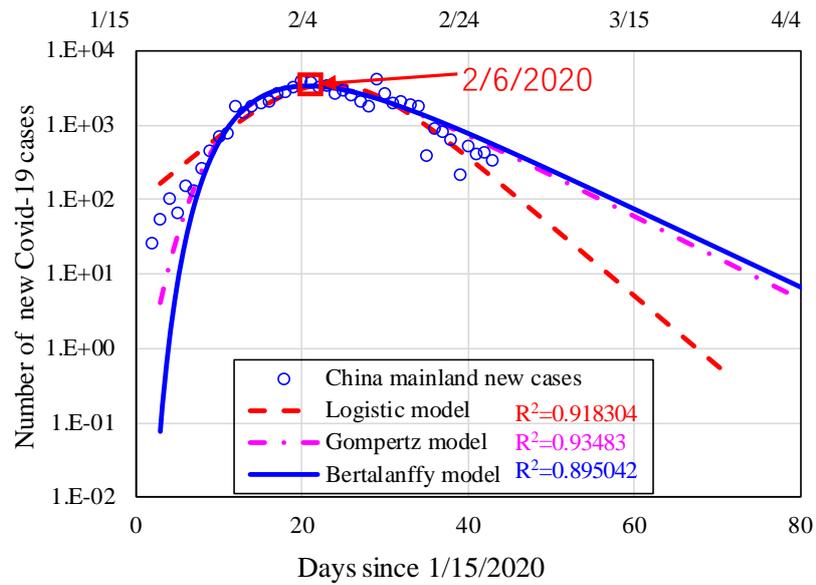

(b) Three Models for Predicting new COVID-19 Cases in China mainland since January 15,2020

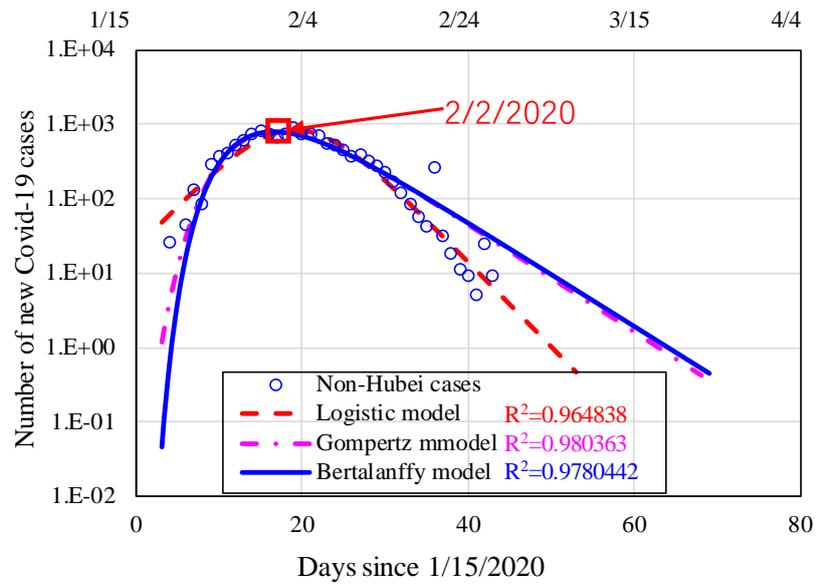

(c) Three Models for Predicting new COVID-19 Cases in non-Hubei areas since January 15,2020

**Figure 4 The prediction of new COVID-19 cases fitted by Gompertz, Logistic and Bertalanffy models**

**4.2 Death Toll**

The death toll of COVID-19 in China is mainly concentrated in Wuhan, Hubei province, so the trend of the death toll in China are basically the same with Wuhan. Similarly, Gompertz, Logistic and Bertalanffy models were used to predict the final death toll of COVID-19. The results are shown in Figure 5.

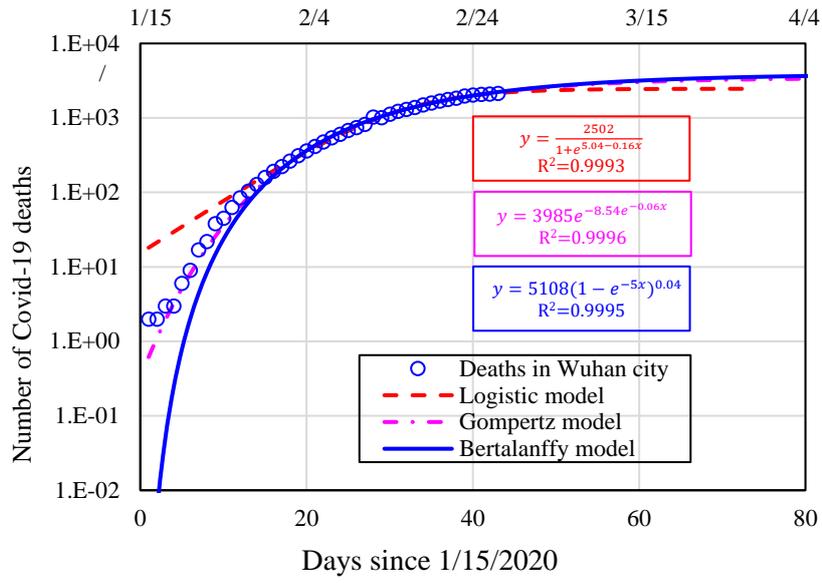

(a) Three Models for Predicting COVID-19 death toll in Wuhan since January 15,2020

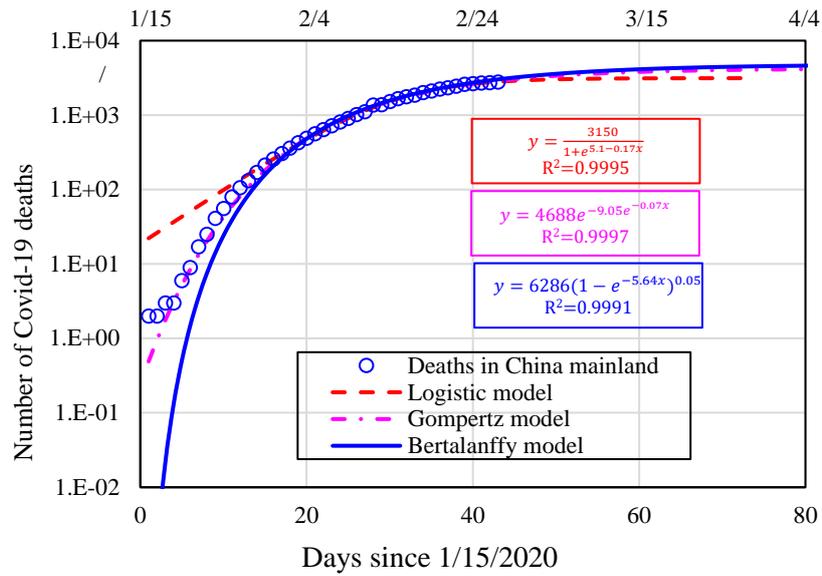

(b) Three Models for Predicting COVID-19 death toll in China mainland since January 15,2020

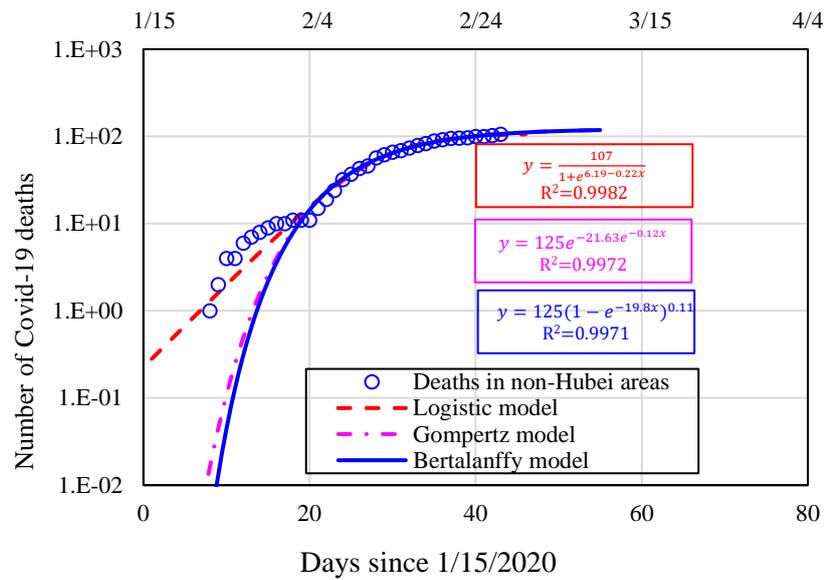

(c) Three Models for Predicting COVID-19 death toll in non-Hubei areas since January 15,2020

**Figure 5 Three Models for Predicting COVID-19 death toll**

The fitting parameters of each model can be seen in Table 2 in which a is the prediction of the death toll; b and c are fitting coefficients; t is the number of days since the first case. $R^2$(DC) means the fitting goodness of cumulative deaths.

According to the available data, the death toll predicted by the three models is Wuhan: 2502-5108; non-Hubei areas: 107-125; China mainland: 3150-6286.As we can see, the results of the death toll predicted by the three models are quite different. It may be due to the fact that the factors affecting the death rate during the epidemic period are more than the cumulative confirmed number and the newly confirmed number, such as the continuous improvement of treatment level, emergency equipment and measures, etc. However, judging from the fitting precision of the models in Figure 4, the Logistic model is obviously better than the other two models. Considering that the later fitting results of the mathematical model is more important than the earlier fitting results, the fitting results of Logistic model may be more accurate, that is, the total death toll of COVID-19 is about 2502 in Wuhan, 107 in non-Hubei areas and 2150 in China mainland, respectively.

**Table 2 Logistic Model, Bertalanffy Model and Gompertz Model for predicting the COVID-19 death toll**

| Model | Parameter | Wuhan | China mainland | Non-Hubei areas |
|---|---|---|---|---|
| Logistic model | a | 2502 | 3150 | 107 |
| | b | 5.04 | 5.1 | 6.19 |
| | c | 0.16 | 0.17 | 0.22 |
| | Cumulative death toll | 2502 | 3150 | 107 |
| | $R^2$(DC) | 0.9993 | 0.9995 | 0.9982 |
| Gompertz model | a | 3985 | 4688 | 125 |
| | b | 8.54 | 9.05 | 21.63 |
| | c | 0.06 | 0.07 | 0.12 |
| | Cumulative death toll | 3985 | 4688 | 125 |
| | $R^2$(DC) | 0.9996 | 0.9997 | 0.9972 |
| | $R^2$(DN) | 0.7529 | 0.8178 | 0.7679 |
| Bertalanffy model | b | 4.99 | 5.64 | 19.8 |
| | c | 0.04 | 0.05 | 0.11 |
| | Cumulative death toll | 5108 | 6286 | 125 |
| | $R^2$(DC) | 0.9995 | 0.9991 | 0.9971 |

## 5 Discussion

The prediction methods of Logistic model, Gompertz model and Bertalanffy model are similar, but the mathematical models are different. From the results, for the prediction of the cumulative number of confirmed diagnoses, the three models can better predict the development trend of the COVID-19 epidemic in the later stages of the epidemic. Among them, the Logistic model is better than the other two models in fitting all the data in Wuhan, while Gompertz model is better in fitting the data in non-Hubei areas. For the prediction of newly confirmed cases, the three models can all well predict the epidemic situation of the COVID-19 in the early and late stages of the epidemic. Among them, the fitting result of Logistic model for all data in Wuhan and non-Hubei areas is better than the other two models. For the prediction of the cumulative death toll, the fitting coefficients of the three models are relatively high, and the figure can be well predicted at the later stage of the epidemic. Various medical resources are becoming more abundant in the later period, the capabilities of medical personnel in various aspects are getting stronger, the support for various resources across the country is getting stronger, and the ability to refine management and treatment is getting stronger. These factors are likely to rapidly reduce the mortality rate of COVID-19. The above factors may also have some influence on the cumulative number of confirmed cases, but due to the large number of confirmed cases, the influence of these favorable human factors on the cumulative number of confirmed cases may be small.

At present, there are only a few papers on the prediction of COVID-19 epidemic. We have collected some COVID-19 epidemic predictions of other researchers, as shown in Table 3. It can be seen from Table 3 that the total prediction results of different models are quite different. According to the prediction results of this article, the cumulative number of confirmed cases will reach maximum in Wuhan and the country around the end of March 2020 at the earliest and around April 2020 at the latest. The results are basically consistent with the results of the Zhong's team[22] that the basic control of the epidemic was at the end of April. The total number of confirmed diagnoses is predicted to be 49852-57447 in Wuhan, 12972-13405 in Non-Hubei areas, and 80261-85140 in China mainland. The predicted total death toll is 2502-5108 in Wuhan, 107-125 in non-Hubei areas, and 3150-6286 in China mainland.

It should be noted that the China mainland data in this article are the data of 31 provinces (autonomous regions, municipalities) and Xinjiang Production and Construction Corps.

In addition, another concerned question is: When will the epidemic of the new coronavirus COVID-19 end? Judging from the SARS situation in 2003, the date corresponding to the maximum number of cumulative diagnoses was basically the date when the epidemic ended. According to the results and data of this article, it is estimated that the epidemic of COVID-19 in novel coronavirus will end at the end of April 2020 in Wuhan and at the end of March 2020 in Non-Hubei areas.

It is worth noting that the above results and conclusions are under the precondition that the prevention and control measures for the epidemic situation of COVID-19 are stable and reliable, foreign cases are not imported into China on a large scale, and the virus of COVID-19 does not produce new and serious acute variations.

Table 3. Results of COVID-19 epidemic by various models (in 2020)

| Model | Inflection point | | | Predicted cumulative number of confirmed cases | | | Predicted date when the cumulative number of confirmed cases reach the maximum | | | Death toll | | |
|---|---|---|---|---|---|---|---|---|---|---|---|---|
| | China | Wuhan | Non-Hubei areas | China | Wuhan | Non-Hubei areas | China | Wuhan | Non-Hubei areas | China | Wuhan | Non-Hubei areas |
| Logistic model | 6 February | 9 February | 2 February | 80261 | 49852 | 12972 | 26 March | 27 March | 8 March | 3150 | 2502 | 107 |
| Gompertz model | 6 February | 9 February | 2 February | 83656 | 56258 | 13328 | 22 April | 24 April | 23 March | 4688 | 3985 | 125 |
| Bertalanffy model | 6 February | 9 February | 2 February | 85140 | 57447 | 13405 | 27 April | 27 April | 24 March | 6286 | 5108 | 125 |
| SEIR model[23] | 23 February | / | / | 55869-84520 | / | / | / | 19 February | / | 2279-3318 | / | / |
| AI Dynamic Model [24] | 16 February | / | / | 42000-60000 | / | / | / | 12-19 February | / | / | / | / |
| SEIR infection model [25] | 1 February | / | / | / | / | 7000 | / | / | / | / | / | / |
| HiddenMarkov model and MCMC method [26] | / | / | / | / | / | / | / | / | / | / | / | / |
| SEIR model based on System dynamics[27] | Before February 9 | / | / | / | / | / | / | 9 February | / | / | / | / |

# 6 Conclusion

(1) It is estimated that COVID-19 will be over probably in late-April, 2020 in Wuhan and before late-March, 2020 in other areas respectively;
(2) The cumulative number of confirmed COVID-19 cases is 49852-57447 in Wuhan, 12972-13405 in non-Hubei areas and 80261-85140 in China mainland;
(3) According to the current trend, the cumulative death toll predicted by the three models are: 2502-5108 in Wuhan,
107-125 in non-Hubei areas, and 3150-6286 in China mainland;
(4) According to the fitting analysis of the existing data by the three mathematical models, the inflection points of the COVID-19 epidemic in Wuhan, non-Hubei areas and China mainland is basically in the middle of February 2020;
(5) The prediction results of three different mathematical models are different for different parameters and in different regions. In general, the fitting effect of Logistic model may be the best among the three models studied in this paper, while the fitting effect of Gompertz model may be better than Bertalanffy model.